\documentclass[aps,superscriptaddress,twocolumn,pre]{revtex4-1}

\usepackage{graphicx,latexsym,amsmath,subfigure,color}

\begin{document}
\title{Critical Casimir Forces and Colloidal Phase Transitions in a Near-Critical Solvent: A Simple Model Reveals a Rich Phase Diagram.}
\setlength{\belowcaptionskip}{-10pt}

\author{John R. Edison}
\affiliation{Soft Condensed Matter, Utrecht University, Princetonplein 5, 3584 CC Utrecht, The Netherlands}
\author{Nikos Tasios}
\affiliation{Soft Condensed Matter, Utrecht University, Princetonplein 5, 3584 CC Utrecht, The Netherlands}
\author{Simone Belli}
\affiliation{Institute for Theoretical Physics, Utrecht University, Leuvenlaan 4, 3584 CE Utrecht, The Netherlands}
\author{Robert Evans}
\affiliation{H.H. Wills Physics Laboratory, University of Bristol, Bristol BS8 1TL, United Kingdom}
\author{Ren\'{e} van Roij}
\affiliation{Institute for Theoretical Physics, Utrecht University, Leuvenlaan 4, 3584 CE Utrecht, The Netherlands}
\author{Marjolein Dijkstra}
\email{m.dijkstra1@uu.nl}
\affiliation{Soft Condensed Matter, Utrecht University, Princetonplein 5, 3584 CC Utrecht, The Netherlands}


\begin{abstract}
From experimental studies it is well-known that colloidal particles suspended in a near-critical binary solvent exhibit interesting aggregation phenomena, often associated with colloidal phase transitions, and assumed to be driven by long-ranged solvent-mediated (SM) interactions (critical Casimir forces), set by the (diverging) correlation length of the solvent. We present the first simulation and theoretical study of an explicit model of a ternary mixture that mimics this situation. Both the effective SM pair interactions and the \textit{full} ternary phase diagram are determined for Brownian discs  suspended in an explicit two-dimensional \textit{supercritical} binary liquid mixture. Gas-liquid and fluid-solid transitions are observed in a region that extends well-away from criticality of the solvent reservoir. We discuss to what extent an effective pair-potential description can account for the phase behavior we observe. Our study provides a fresh perspective on how proximity to the critical point of the solvent reservoir might influence colloidal self-assembly.
\end{abstract}

\maketitle

Colloidal particles dispersed in a binary solvent mixture have an inherent preference for one of the two solvent species. This is reflected by preferential adsorption of the favoured species on the colloid surface, leading to the development of adsorbed films. Such films can mediate an effective interaction between two colloidal particles which is remarkably sensitive to the thermodynamic state of the solvent. Close to the (demixing) critical point of the solvent the adsorbed film thickness is determined by the correlation length $\xi$ of the solvent \cite{Floeter1995} and, as first predicted by Fisher and de Gennes \cite{FISHER1978}, the resulting solvent-mediated (SM) interactions are long-ranged, with universal scaling properties. An analogy between the confinement of quantum fluctuations of the electromagnetic field \cite{Casimir1948} and that of thermal composition fluctuations in a near-critical binary solvent led to these (universal) SM forces being referred to as critical Casimir forces \cite{Krech1994}. 

Theoretical studies on near-critical fluids confined between a pair of infinitely large planar walls (representing two static large colloids) \cite{Krech1994,Evans1994,Hanke1998,Krech1999,Vasilyev2009}, along with direct experimental measurements of the Casimir force \cite{Hertlein2008,Gambassi2009}  between a colloid and a wall  have advanced our understanding of the nature of two-body SM interactions. Although experimental investigations of a \textit{suspension} of colloids go back to the pioneering work of Beysens and Esteve \cite{Beysens1985}, for a very recent experimental study see \cite{Nguyen2013}, the theory and computer simulation of such systems remain at a primitive stage. Here we use computer simulations of a simple model to understand the strength and range of the SM interactions and the resulting phase behaviour of a dense colloidal suspension as a function of the thermodynamic state of the solvent. Computer simulation of colloids in an explicit molecular solvent with a bulk correlation length that diverges upon approaching the critical point is notoriously difficult as very different length and time scales are involved. Nevertheless, by sacrificing one spatial dimension and using a lattice model, we have calculated the  phase diagrams for an explicit ternary  solvent-solvent-colloid mixture, without resorting to the assumption of pairwise effective potentials employed in other studies, most notably \cite{Mohry2012,Dang2013,Mohry2014}.

\begin{figure}[h]
\centering
\includegraphics[width=2in,keepaspectratio]{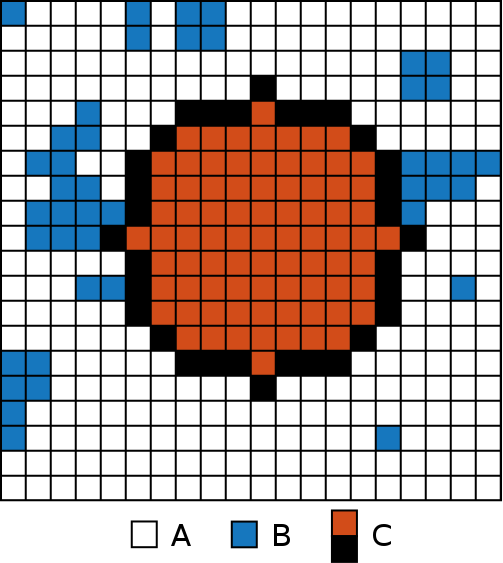}
\caption{A schematic representation of the solvent-solvent-colloid lattice model. White cells are occupied by solvent species A, blue cells by solvent species B, and brown and black cells represent the interior and the boundary of a single colloidal particle C, respectively.} 
\label{fig01}
\end{figure}

Following Rabani \textit{et al.} \cite{Rabani2003} we model the ternary solvent-solvent-colloid mixture as an incompressible ABC mixture on a 2D square lattice, as shown schematically in Fig. \ref{fig01}. Colloids C are discretized hard discs with a radius of $R$ lattice sites, occupying a fraction $\eta$ of the lattice sites. Every site that is left unoccupied by the colloidal discs is occupied by either a solvent molecule of species A or B, such that the fraction of sites occupied by A and B equals $1-\eta-x$ and $x$, respectively. We consider only nearest neighbour AB repulsions and BC attractions: an energy penalty $\frac{1}{2}\epsilon>0$ is assigned to every nearest neighbour AB pair to drive AB demixing at sufficiently low temperatures $T$, and an energy gain $-\frac{1}{2}\alpha \epsilon$ with $\alpha \geq 0$ for every BC pair to mimic the colloid C's preference for species B. Throughout we set the lattice spacing to unity. We investigated carefully several lattice effects present in our model and we discuss it briefly in section V of the supplementary information (SI). We note that lattice effects have no implications for the key results of our work.

\begin{figure}
\centering
\subfigure{\includegraphics[width=2.250in]{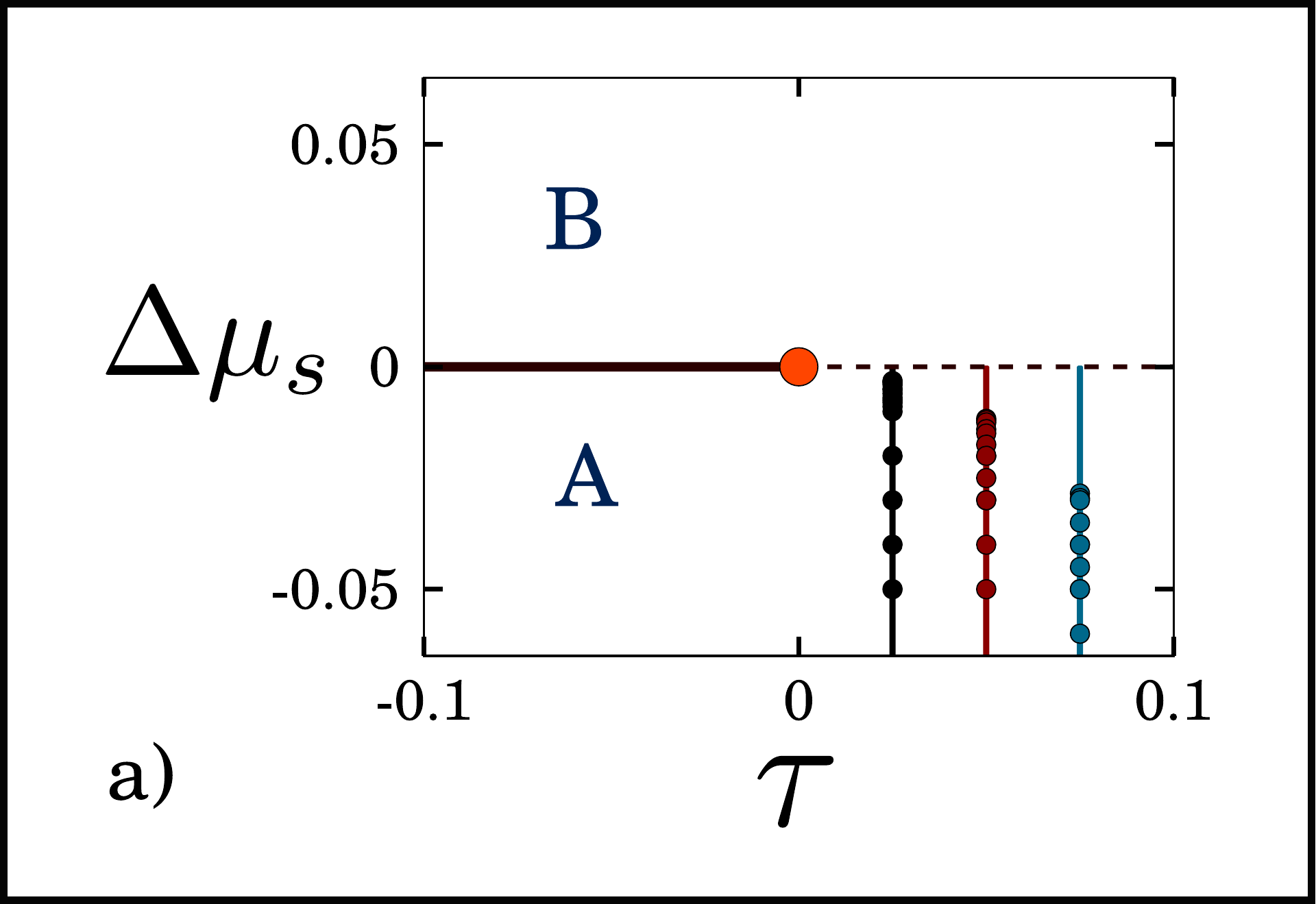}\label{fig02a}}
\subfigure{\includegraphics[width=2.250in]{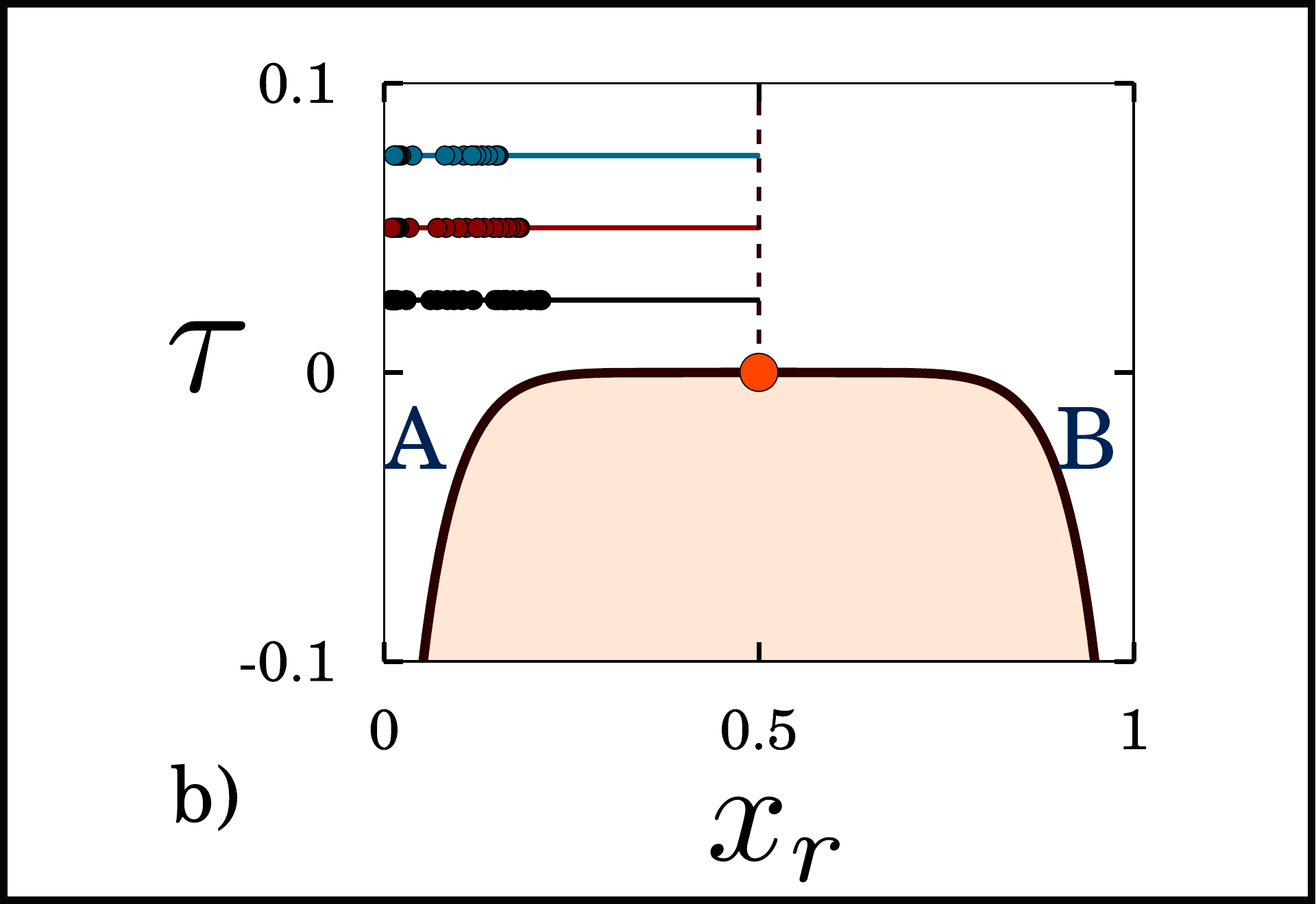}\label{fig02b}}
\caption{a) Phase diagram of the colloid-free AB solvent mixture plotted as $\Delta \mu_s$ vs $\tau$. b) Phase diagram (binodal) of the same mixture plotted as  $\tau$ vs $x_r$, the composition of the solvent. The two coexisting phases for $\tau<0$ are designated A and B. In a) and b) the lines correspond to the paths along which the phase diagram of the full ternary ABC mixture is determined; the dots represent the states where phase coexistence is observed and the orange dot indicates the critical point of the AB solvent mixture $(\tau = 0, \Delta \mu_s = 0)$.} 
\label{fig02}
\end{figure}
 
In the limit $\eta = 0$, our model reduces to a binary AB mixture that is isomorphic to the 2D lattice (Ising) model. The critical temperature of this binary mixture is $T_c=0.567\epsilon/k_B$, and its thermodynamic state is characterized fully by the reduced temperature $\tau=(T-T_c)/T_c$ together with either the reduced chemical potential difference $\Delta\mu_s=(\mu_B - \mu_A)/\epsilon$ between species B and A or the composition $x$.  For $\Delta \mu_s <0$ the AB mixture favours an A-rich composition at all temperatures. Moreover for $\tau<0$ demixing into an A-rich state  ($x<0.5$) and a B-rich state ($x>0.5)$ takes place at $\Delta\mu_s=0$, with a critical point $\{\tau_c=0,x_c=0.5\}$, see Fig. 2a and b.  In the two limits $\Delta\mu_s \rightarrow\pm\infty$ our ABC mixture reduces to the 2D AC or BC hard-disc system with packing fraction $\eta$ in an (irrelevant) pure A solvent ($x=0$) or pure B solvent ($x=1-\eta$).  Barring small discretization and lattice artifacts, and ignoring subtleties regarding the (non-)existence of a stable hexatic phase, these AC and BC systems exhibit fluid-solid coexistence for $\eta\in[0.700,0.716]$ as represented by vertical dashed lines in Fig. 3a-c \cite{Bernard2011}.

Throughout this work we study colloids immersed in a \textit{supercritical} (one-phase) AB mixture, relatively poor in the colloid-preferred species B ($ \eta > 0, \tau > 0$, and $\Delta \mu_s\leq 0$). This choice precludes solvent-mediated colloidal aggregation arising from complete wetting and capillary condensation \cite{Evans1990}. The solvent is treated grand canonically, i.e. the system is in thermal and diffusive contact with an AB solvent reservoir with composition $x_r$ that fixes $\tau$ and $\Delta \mu_s$. The ABC mixture has composition $x \neq x_r$. Note $\tau$ merely sets the temperature; it is not a measure of distance from criticality of the ternary mixture. 

\begin{figure*}
\centering
\includegraphics[width=5.50in,keepaspectratio]{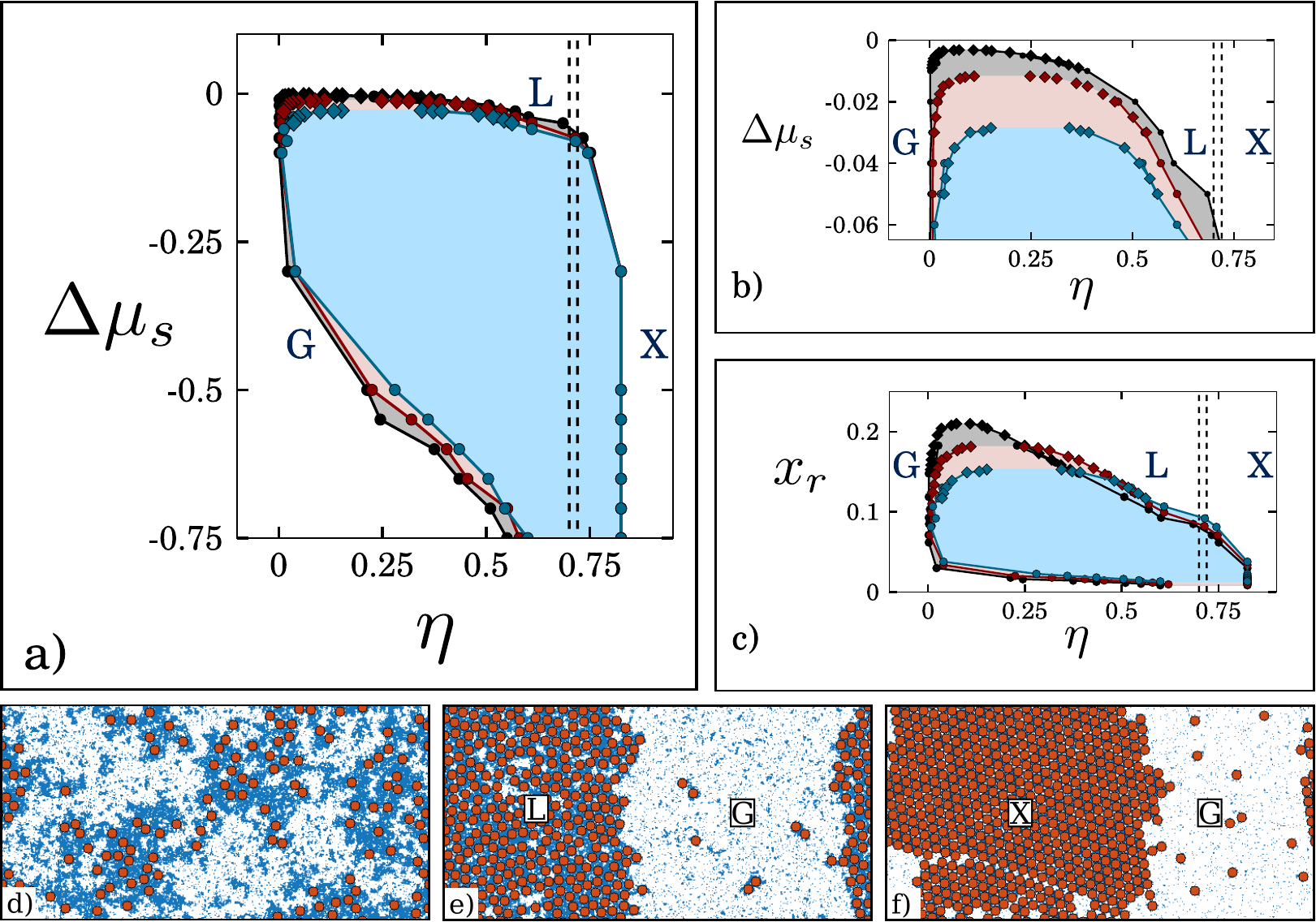}
\caption{ a) Phase diagrams of the full ABC ternary mixture plotted as solvent chemical potential $\Delta \mu_s$ vs hard disc (colloid) packing fraction $\eta$ for $R=6$. b) The top portion of a) is replotted for clarity. c) Phase diagram of the ABC model plotted as reservoir solvent composition $x_r$ vs $\eta$. Black, red and blue symbols refer to $\tau = 0.025$, $0.05$ and $0.075$. The diamonds and dots denote the phase boundaries obtained from grand canonical staged insertion Monte Carlo (MC) simulations and $( \eta, \tau, \Delta \mu_s )$ - ensemble MC simulations, respectively. The vertical dashed lines denote fluid-solid coexistence for pure hard discs. d - f ) Simulation snapshots of a system of $256\times 512$ lattice sites at $\tau=0.05$ and $\alpha = 0.6$, showing d) a supercritical colloidal phase at $\Delta \mu_s = -0.005$ of 128 colloids, e) gas-liquid (G-L) coexistence at $\Delta \mu_s = -0.04$ of 348 colloids, and f) gas-crystal (G-X) coexistence at $\Delta \mu_s = -0.3$ of 580 colloids.}
\label{fig03}
\end{figure*}

We focus on the case where B-rich layers adsorbed on the colloid surfaces compete with a supercritical A-rich bulk solvent ($\tau>0$ and $\Delta\mu_s<0$, $\alpha=0.6$). We perform simulations of the model in the fixed $(\eta,\tau, \Delta \mu_s)$-ensemble (see section I of the SI) and in the grand ensemble, using the staged insertion method \cite{Ashton2011} together with the Transition Matrix Monte Carlo (TMMC) technique \cite{Errington2003}, to accurately determine phase coexistence. The GC-TMMC results reported here are for a system size $L=256$. For a few state points we compared the results for two different system sizes, $L=256$ and $L=512$, and found the coexisting packing fractions to be the same up to the third decimal place. In Figs. \ref{fig03}a and b we present the phase diagram  of the ternary mixture in the $\Delta \mu_s$ vs $\eta$ representation for $\tau = 0.025, 0.05, 0.075$. The correlation length of the AB solvent reservoir at the isochoric composition ($\Delta \mu_s = 0, x_r=0.5$) is $\xi = 0.567/\tau$, and for these temperatures it is comparable to the size of the colloid $R=6$. Although the underlying AB solvent  reservoir  is supercritical, our simulations reveal that a non-zero concentration of large Brownian discs induces stable colloidal gas (G), liquid (L), and crystal (X) phases as well as two-phase G-L and G-X coexistence.  The G-L coexistence, shown more clearly in Fig. \ref{fig03}b, terminates at a critical point that shifts to lower $\Delta \mu_s$ and higher $\eta$ with increasing $\tau$. Tracing the locus of the three critical points of the ABC mixture from $\tau=0.075$ through $0.05$ to $0.025$, it appears that the critical points of the colloidal G-L transition are continuously connected to that of the binary solvent mixture ( $\eta=\tau=\Delta\mu_s=0$); investigations at smaller $\tau$ are constrained by our computational resources. 

For $\Delta \mu_s<-0.1$ we also observe G-X coexistence with a broad colloid density gap that narrows sharply upon lowering $\Delta \mu_s$, consistent with the limiting hard-disc fluid-solid coexistence at $\Delta \mu_s \rightarrow -\infty$ (vertical dashed lines). Significantly, this decreasing density gap at G-X coexistence suggests an additional underlying metastable G-L lower critical point. Although we have not been able to identify this in our MC simulations, such an additional critical point \textit{does} occur in our mean-field treatment presented in the SI. Moreover, if we accept hard-disc coexistence in the opposite limit $\Delta \mu_s \rightarrow \infty$, then we also expect a G-L-X triple point at $\Delta \mu_s \simeq -0.06$ for $\tau = 0.025$ (see Fig. \ref{fig03}b), and at even lower $\Delta \mu_s$ for higher $\tau$.

In Figs. \ref{fig03}d - \ref{fig03}f and in  SI-movie-03/04/05 we show results for  systems of $256 \times 512$ lattice sites simulated at reduced temperature $\tau=0.05$ ($\alpha = 0.6, R=6$) that illustrate configurations of (d) a supercritical (homogeneous single-phase) fluid state, (e) G-L coexistence, and (f) G-X coexistence. In all three cases the local solvent composition is strongly correlated with the local colloid density, such that the coexisting L phase in (e) and X phase in (f) have a binary BC composition with tiny traces of A. Conversely in the coexisting G phases shown in Figs. \ref{fig03}e and \ref{fig03}f the solvent composition is very close to the composition of the reservoir $x \simeq x_r$. In Fig. \ref{fig03}c we convert  the phase diagram of Fig. \ref{fig03}a into the $x_r-\eta$ representation. It is evident from the snapshots and Figs. \ref{fig03}c and \ref{fig02}b that for all observed G-L and G-X coexistence: (i) the composition of the solvent reservoir  $x_r<0.25$ is far from its critical composition $x_c=0.5$, and (ii)  the correlation length of the solvent is smaller than the colloid radius, $\xi<R$.  Strikingly, in the homogeneous supercritical state of Fig. 3d the correlation length (the  typical size of the A-rich and BC-rich  "patches") is clearly much larger than the colloid radius and thus far exceeds that of the solvent reservoir. This reflects the nearby (G-L) critical point of the ternary mixture which, as noted previously, appears to be continuously connected to the critical point of the binary AB solvent mixture ($\eta=\tau=\Delta\mu_s=0$). In fluid mixtures where the species interact via short-range potentials, all structural correlations decay with the same correlation length \cite{Evans1994_a}. Therefore along the G-L critical locus, solvent-solvent, colloid-colloid and solvent-colloid correlations should decay with the same, diverging correlation length. We have confirmed this numerically by calculating the BB, BC and CC pair correlation functions; see Fig. S3 of the SI. We have also confirmed the divergence of the long wavelength limit of the structure factor $S_{BB}$ upon approaching the critical point of the ternary mixture at a fixed value of temperature $\tau=0.025$; see Fig. S4 of the SI. 

\begin{figure}[h]
\centering
\includegraphics[width=3in]{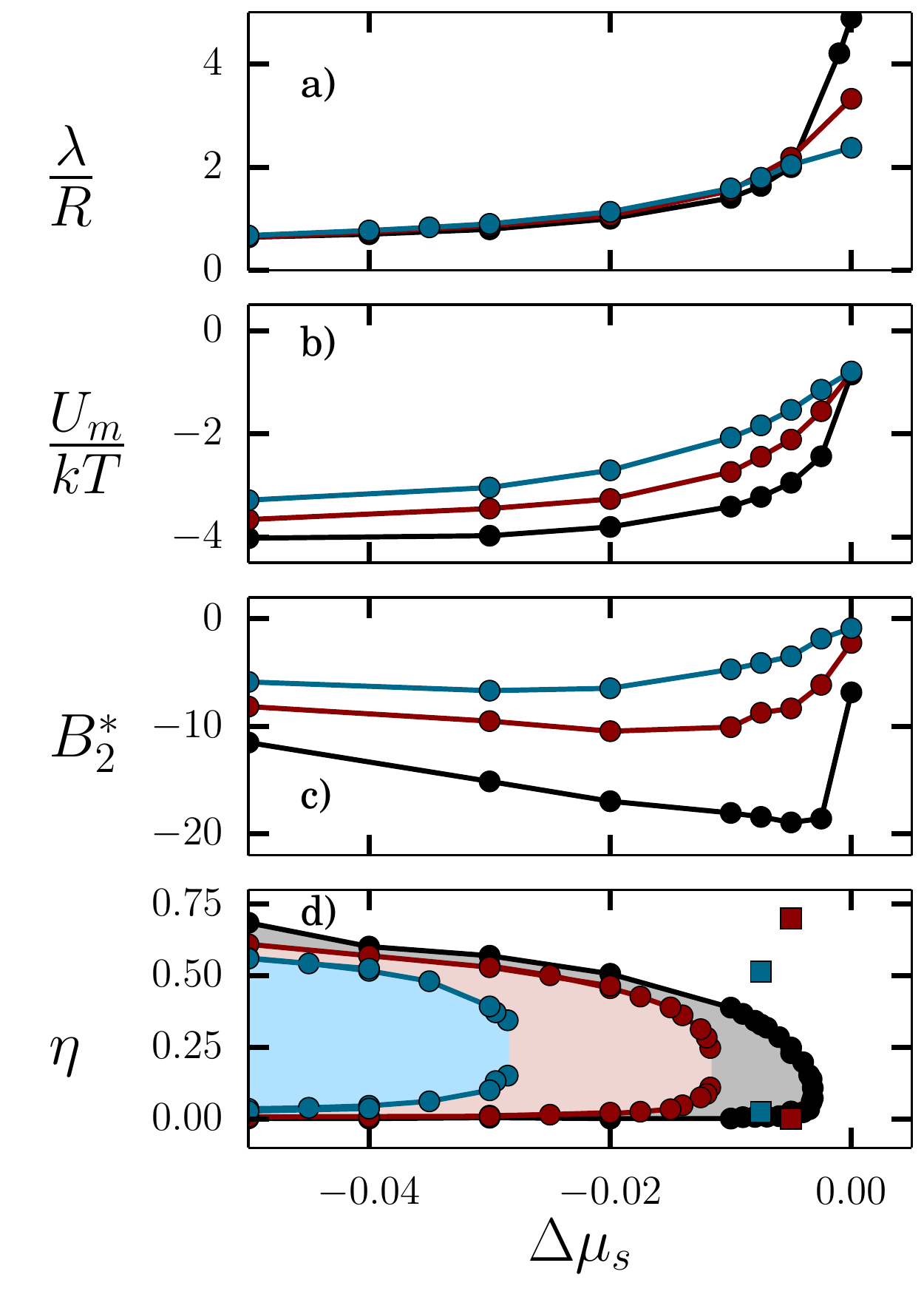}
\caption{a) Thickness $\lambda$ of the B-rich film adsorbed on a single colloid, b) The minimum well depth of the effective two-body SM potential, c) The reduced second virial coefficient $B_2^*$, d) To facilitate comparison, the phase boundaries of the ternary ABC mixture of Fig. 3 are replotted in the $\eta-\Delta \mu_s$ representation. The square symbols (Fig 4d ) correspond to packing fractions of coexisting phases computed from simulations of colloids interacting via the measured two-body interaction. Black, red and blue symbols refer to reduced temperatures $\tau = 0.025$, $0.05$ and $0.075$ respectively  ($R=6$, $\alpha=0.6$).}
\label{fig04}
\end{figure}

As mentioned earlier, there have been several attempts to ascertain the phase behaviour of colloids in a near-critical solvent based solely on effective two-body interactions, e.g. \cite{Mohry2012, Nguyen2013}. In order to assess the validity of this approach for the present model system  we calculated several one- and two-colloid properties for the range of thermodynamic state points studied above. For the three temperatures investigated we show in Fig. \ref{fig04} the dependence on $\Delta \mu_s$  of (a) the thickness $\lambda$ of the adsorbed B-rich film on a single disc, (b) the minimum $U_{m}$ of the effective pair potential $U(x,y)$. (c) the  reduced second virial coefficient $B_2^*=  (1/2) \int_{0}^{\infty}  \int_{0}^{\infty} (1 - \exp[-U(x,y)/k_BT]) dx dy / 2 \pi R^2$, normalized to that of 2D  hard discs.  The effective two-body potential $U(x,y)$ is obtained as follows. We simulate a system of just two colloids suspended in solvent at fixed $\{\Delta \mu_s, \tau\}$. We fix the position of one colloid fixed at $(0,0)$, and compute the probability $P(x,y)$ of finding the other colloid at position $(x, y)$, using the TMMC technique (see section V of the SI for details). To facilitate comparison, we replot the phase diagrams of Fig. 3 in the $\eta-\Delta \mu_s$ representation in Fig. 4d. The film thickness $\lambda$ and the well depth $U_{min}$ are measures of the range and strength of $U(x,y)$, respectively. The quantity $B_2^*$ is a well-established (dimensionless) measure of combined strength and range. This must be sufficiently negative in order for gas-to-liquid condensation to occur in systems described by pairwise additive interactions \cite{Noro2000, Vliegenthart2000}. 

Fig. \ref{fig04}a shows a monotonic increase of the film thickness from $\lambda\ll R$ to  $\lambda\gg R$,  reflecting the growth of the correlation length, as the isochoric composition is approached ($\Delta \mu_s \rightarrow 0$). In the same range $U_{m}$ varies non-monotonically being strongest at slightly negative $\Delta \mu_s$, reaffirming earlier theoretical predictions \cite{Drzewinski2000,Okamoto2012}. At $\Delta \mu_s \simeq 0$ the effective pair potential is long-ranged, however it is only \textit{weakly} attractive ($|U_{m}| / k_B T < 1$). Upon decreasing $\Delta \mu_s$,  $U(x,y)$ does become more attractive, although the adsorbed film thickness $\lambda$ and thereby the range of $U(x,y)$ decreases. $B_2^*$ also becomes more negative. 

We performed simulations of the \textit{effective} system, with the colloid-colloid pair interaction determined by the measured effective potential. For states where the actual ternary mixture is super-critical, e.g $\{\tau = 0.075$, $\Delta \mu_s = -0.0075\}, \{\tau = 0.05$, $\Delta \mu_s = -0.005\} $ (c.f Figs. \ref{fig03}a, \ref{fig04}d), simulations performed with the effective two-body potential predict G-L coexistence. The square symbols in Fig. \ref{fig04}d denote the packing fractions of coexisting phases at these two representative state points. This along with the $B_2^*$ and $U_m/kT$ curves indicates that the approaches employing only effective pair potentials as obtained from e.g. planar slit studies and the Derjaguin approximation overestimate the extent of G-L coexistence, and underestimate the shift in critical point of the ternary mixture with respect to that of the solvent reservoir. 

In summary, we find the phase behaviour of a model of colloids in a near-critical solvent to be rich; we observe (i) G-L and G-X coexistence with accompanying solvent demixing, (ii) both occur far from the critical point of the solvent reservoir and the locus of G-L critical points appears to connect smoothly to this and (iii) many-body interactions are crucial to account quantitatively for the observed colloidal phase behaviour.  In light of our results it would be interesting to revisit  the problem of protein assembly in two-dimensional plasma membranes of living cells \cite{Veatch2008, Machta2012} and the recent experiments of Nguyen et al\cite{Nguyen2013}. The topology of the phase diagram of colloidal particles in a near-critical binary solvent stems from an intricate balance between competing colloid-solvent and solvent-solvent couplings that can only be captured properly in a treatment of the full ternary mixture. Moreover, we speculate that the topology is likely to hold for an analogous 3D system (hard-sphere colloids); there is nothing particular to two dimensions, an assertion supported by our mean-field treatment - see SI.


\footnotesize
\begin{acknowledgments}
We thank N. Wilding, D. Ashton and A. Macio{\l}ek for stimulating discussions. J.R.E. and M.D. acknowledge financial support from a Nederlandse Organisatie voor Wetenschappelijk Onderzoek (NWO) VICI grant. N.T. and M.D.acknowledge financial support from an NWO-ECHO grant. J.R.E., N.T. and M.D acknowledge a NWO-EW grant for computing time in the Dutch supercomputer Cartesius. R.E. acknowledges financial support from the Leverhulme Trust. \\ 
\end{acknowledgments}
\pagebreak
\section{Supplementary information}

\begin{figure*}
\centering
\includegraphics[width=7.5in]{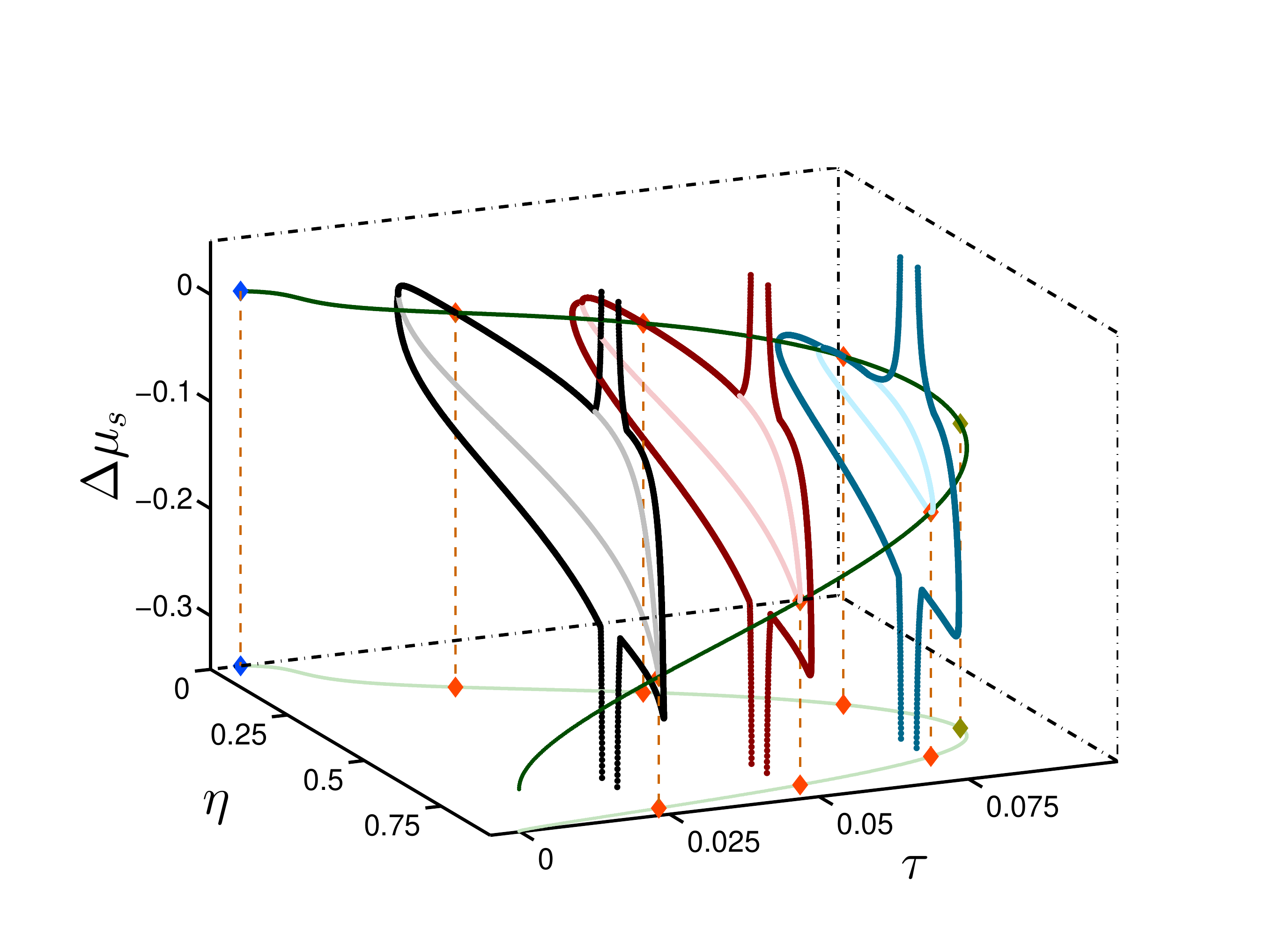}
\caption{\label{SI_fig01} \textbf{Phase behavior of the ternary mixture as predicted by mean-field theory}: Binodals of the ternary colloid solvent system as calculated within mean-field theory plotted in the $\Delta \mu_s$ vs $\eta$ vs $\tau$  representation. We show slices of the full phase diagram for three fixed temperatures $\tau = 0.025$ (black), $\tau = 0.05$ (dark red ), and $\tau = 0.075$ (blue). The gray, pale red and pale blue curves correspond to metastable colloidal gas-liquid coexistence, which also terminates at a critical point. The dark green curve is the locus of critical points of the ternary mixture, this approaches smoothly the critical point of the solvent denoted by the blue diamond dot in the limit $\tau \rightarrow \tau_c^{MF} = 0.0$, $\eta = 0$, $\Delta \mu_s = 0$. For each $\tau$ we show the upper (stable) and lower (metastable) G-L critical points as indicated by the orange diamond symbols. The olive green diamond symbol corresponds to the point where the upper and lower critical points of the ternary system merge and disappear. The dashed orange lines (guide to the eye) connect the critical points to their projection in the $\eta-\tau$ plane. The projection of the locus of critical points in the $\eta-\tau$ plane is given by the pale green curve.}
\end{figure*}

\begin{figure*}
\centering
\subfigure[]{\includegraphics[width=3in]{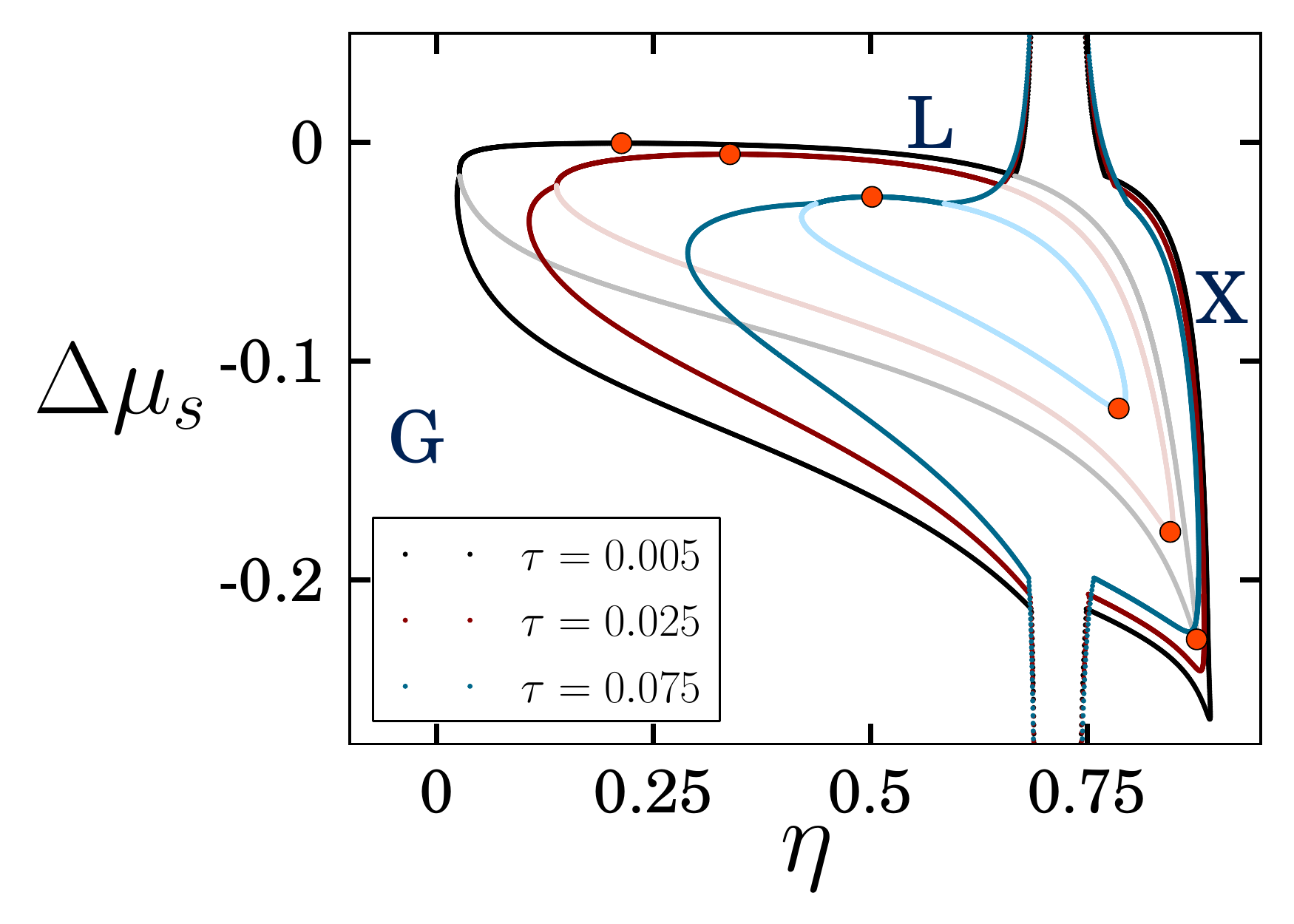}}
\subfigure[]{\includegraphics[width=3in]{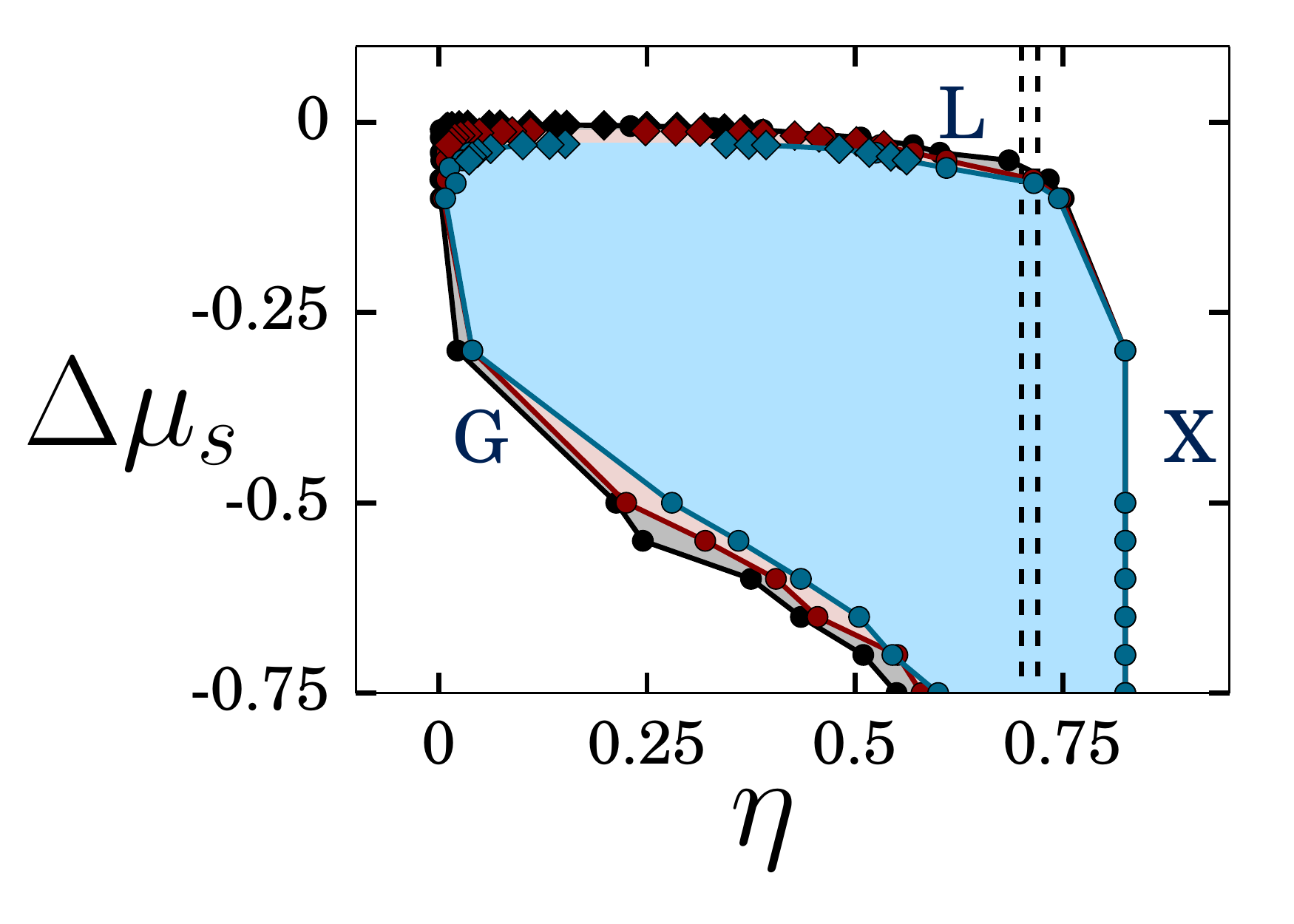}}
\caption{\label{SI_fig02} a) Projections of the binodals of the ternary colloid-solvent system as calculated within mean-field theory for three fixed temperatures $\tau = 0.025$ (black), $\tau = 0.05$ (dark red ), and $\tau = 0.075$ (blue) (same as shown in figure \ref{SI_fig01}), on the $\Delta \mu_s$ -  $\eta$ plane. b) Phase diagrams of the ABC model computed with simulations for three fixed temperatures $\tau = 0.025$ (black), $\tau = 0.05$ (dark red ), and $\tau = 0.075$ (blue) (same as Fig. 3a of our manuscript).}
\end{figure*}

\subsection{Simulation Methods}

Our model is based on that of Rabani {\em et al.} \cite{Rabani2003}. We model the colloidal suspension as an incompressible ABC mixture on a 2D square lattice. Colloids C are discretized hard discs (HD) with a radius of $R$ lattice sites that can undergo translational motion on the square lattice. The hard-disc Hamiltonian $H_C$ is zero for non-overlapping configurations, and infinite if any pair of colloids overlap. Every lattice site $i$ has an  occupancy number $n_i=1$ if it is occupied by a  colloidal disc, and 0 if it is available for an A or a B solvent molecule. For sites with $n_i = 0$ we associate an occupancy number $s_i=-1$ if the site is occupied by A, and $s_i=1$ if by B. We consider only nearest neighbour interactions and assign an energy penalty $ \epsilon/2>0$ for every nearest neighbour AB pair  to drive AB demixing at sufficiently low temperatures and an energy gain of $-\alpha \epsilon / 2$ with $\alpha \geq 0$ for every BC pair to mimic preferential adsorption of solvent B on the colloid surfaces.  The total Hamiltonian thus reads

\begin{equation}
\label{eq01}	
H = H_C +\frac{\epsilon}{4}\sum_{\langle i,j\rangle}(1-s_i s_j)(1 - n_{i})(1 -n_{j})-\frac{\alpha \epsilon}{4} \sum_{\langle i,j \rangle}n_i (1+s_j)(1 - n_j)
\end{equation}

where the summation runs over the set of distinct nearest neighbour pairs $ij$, and for every lattice site $i$, $n_i=1$ is it is occupied by a  colloidal disc, and 0 if it is available for an A or a B solvent molecule. For sites with $n_i = 0$ we associate an occupancy number $s_i=-1$ if the site is occupied by A, and $s_i=1$ if by B. 

We performed simulations in an elongated simulation box of $256\times 512$ sites in the fixed $(\eta,\tau, \Delta \mu_s)$-ensemble. For packing fractions  $\eta$ of hard discs that lie within the binodal curve, two-phase coexistence will be observed in the simulation box. The packing fractions of the coexisting phases can be obtained from the resulting density profiles of the hard discs. In order to determine the G-L coexistence more accurately we treat the colloids grand canonically using the staged-insertion technique \cite{Ashton2011} in combination with the transition matrix (TM) MC method, see e.g.\cite{Errington2003,Escobedo2007,Ashton2011}.

The length $L$ of the simulation box in all our simulations is at least 4 times the correlation length of the bulk solvent reservoir at the composition $x_c = 0.5$, (the maximum correlation length of the solvent reservoir at a fixed $\tau$). We have also taken care to simulate for time scales much longer that the slowest correlation time in the system. The GC-TMMC simulation results reported in our manuscript are for a system size $L=256$. A typical GC-TMMC run, to locate one coexistence point for the $L=256$ system, takes $ \simeq 600 $ CPU hours. For a few state points we compared the results for two different system sizes, $L=256$ and $L=512$, and found the coexisting densities to be the same up to the third decimal place. The $L=512$ system required $\simeq 5000$ CPU hours to simulate one state point. It is not feasible to perform TMMC simulations in system sizes larger than $L=512$.

\subsection{Mean Field Phase diagram : 3D Representation}

Within a mean-field approximation we analyzed the Helmholtz free energy associated with the Hamiltonian of our ABC model, which  can be decomposed as $F_{MF}=F_C + F_{AB} + U_{BC}$, with (i) the pure-colloid contribution $F_C(\eta, T)$ (ii) the mean-field free energy $F_{AB}(x,\eta,T)$ of the binary AB mixture in the free space in between the colloids (with fractions $1-x'$ and $x'\equiv x/(1-\eta)$ of A and B, respectively), and (iii) the average adsorption energy $U_{BC}$ of the B solvent on the colloid surfaces.  This yields, up to irrelevant constants, 

\begin{align}
F_{MF}(\eta,T,x) &= {F_{C}(\eta,T)} + \frac{2 \epsilon x(1-x-\eta)}{(1-\eta)} + \nonumber \\ & k_B T \left[ x \ln \frac{x}{1-\eta} + (1 - x - \eta) \ln \left(\frac{1-x-\eta}{1-\eta}\right) \right] \nonumber \\ & -\frac{Z \alpha \epsilon }{v_c}\frac{x \eta}{1-\eta}
\label{eq03}
\end{align}	

where $Z\simeq 2\pi R$ is the effective colloidal coordination number and where $v_c\simeq \pi R^2$ is the effective volume (area in 2D) of the colloid. For $F_C(\eta, T)$ we employ the hard-disc free energy from Ref. \cite{Santos1995} for the fluid phase, and from Ref. \cite{Young1979} for the solid phase. The phase diagram shown in Fig. \ref{SI_fig01} is based on $Z \alpha = 32$ and $v_c=1000$, which do not correspond to values used in our simulation studies. Our objective here is to attempt to understand the topology of the simulation phase diagrams \textit{qualitatively} and investigate the possibility of a lower (metastable) G-L critical point. 

In Fig. \ref{SI_fig01}, we plot the resulting phase diagrams for various $\tau>0$, which reveal a closed-loop immiscibility gap and  \textit{two} G-L critical points. We plot three slices of the full phase diagram and the locus of critical points of the ternary mixture. This is shown as the dark green curve in the figure, and it smoothly approaches the critical point of the pure solvent mixture $\tau_c^{MF} = 0.0$ (blue diamond symbol). The locus of critical points indeed continues for $\tau < \tau_c^{MF}$. For clarity, we do not present this. The pale green curve is the projection of the critical line on the $\eta-\tau$ plane. The line of colloidal G-L critical points in our simulations should also behave in a similar manner. Furthermore the mean-field theory predicts that at a fixed temperature, there exists an upper G-L critical point and a lower metastable G-L critical point. On increasing temperature the two critical points approach each other, merge and disappear at a certain temperature. This point is indicated by the olive green diamond symbol in Fig. \ref{SI_fig01}. We also observe coexistence of two crystal phases with the same (hexagonal) symmetry but different lattice spacings, also terminating at a critical point. The \textit{topology} of the mean-field phase diagram and its $\tau$-dependence are remarkably consistent with that obtained from simulations. 

\subsection{Correlation functions}

\begin{figure}
\centering
\includegraphics[width=3in]{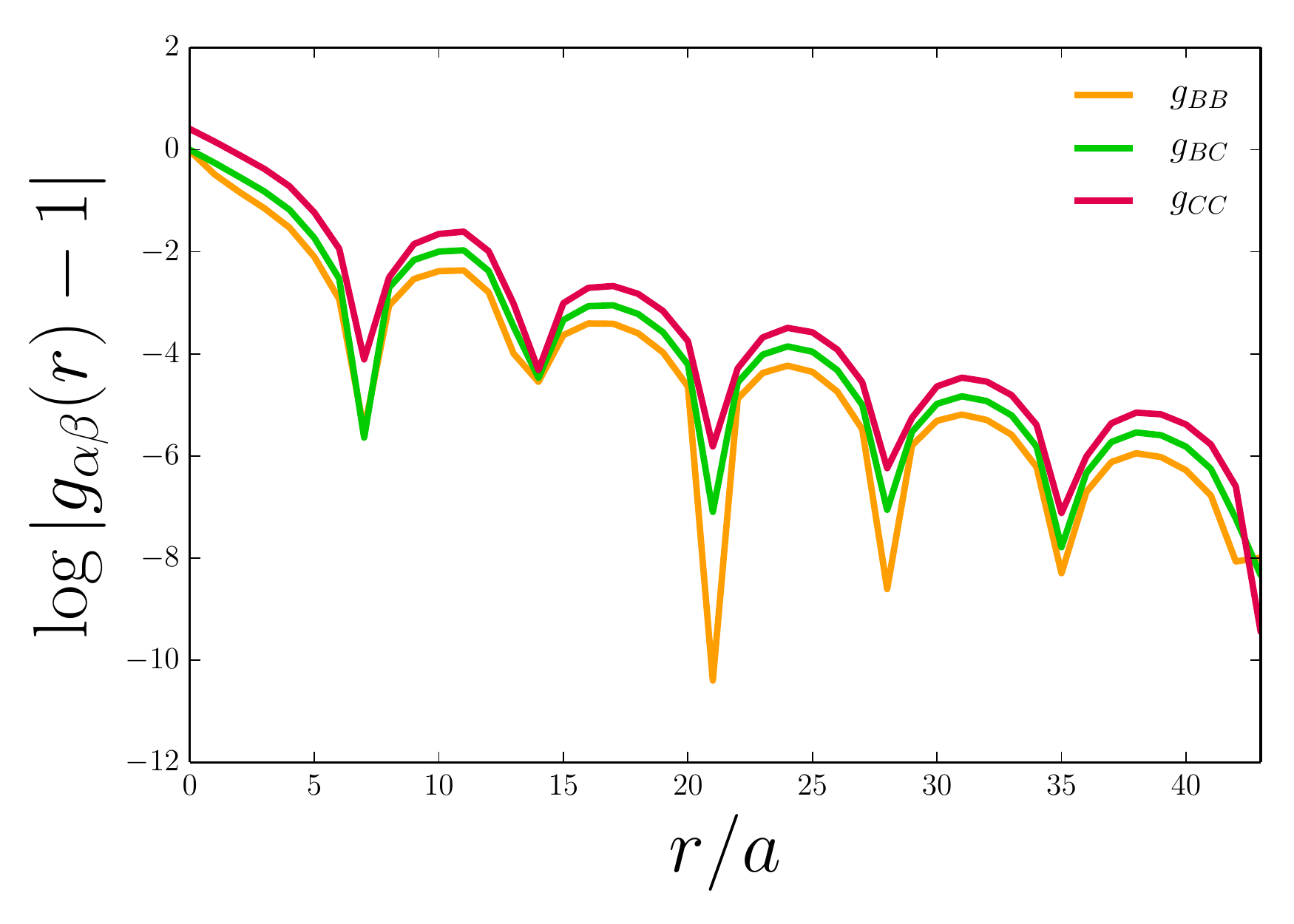}
\caption{\label{SI_fig03} The partial pair correlation functions plotted as $\log \left| g_{\alpha \beta}(r) - 1\right|$ vs distance, normalized by the lattice spacing $a$, at temperature $\tau = 0.025$, colloid packing fraction $\eta = 0.4$ and chemical potential $\Delta \mu_s = -0.005$. All three correlation functions exhibit the same decay length and period, as predicted by \cite{Evans1994_a}.}
\end{figure}

We define the two-point correlation functions as,

\begin{equation}
    g_{\alpha \beta}(\mathbf{r}) = \frac{N}{\left< N_{\alpha} \right> \left< N_{\beta} \right>} \left<\sum_{\left\{i,j | \mathbf{r}_i - \mathbf{r}_j = \mathbf{r}\right\}}n^{\alpha}_i n^{\beta}_j\right>
    \label{eq:2pcon}
\end{equation}

where $n^{\alpha}_i$ represents the occupancy of species $\alpha$ at site $i$, $N_{\alpha}$ is the total number of sites filled with species $\alpha$, and $N = \sum_{\alpha} N_{\alpha}$, is equal to the lattice size. It is now well-established that in fluid mixtures where all species interact via short range potentials, all structural correlations should decay with the same correlation length \cite{Evans1994_a}. In Fig. \ref{SI_fig03}, we plot the quantity $\log \left| g_{\alpha \beta}(r) - 1\right|$, for the pairs BB, BC and CC. It is evident that all correlations do decay with the same correlation length, as expected \cite{Evans1994_a}, illustrating that the correlations of all species of our ternary mixture remain coupled.

\subsection{Structure factors}

\begin{figure}
\centering
\subfigure[]{\includegraphics[width=3.0in]{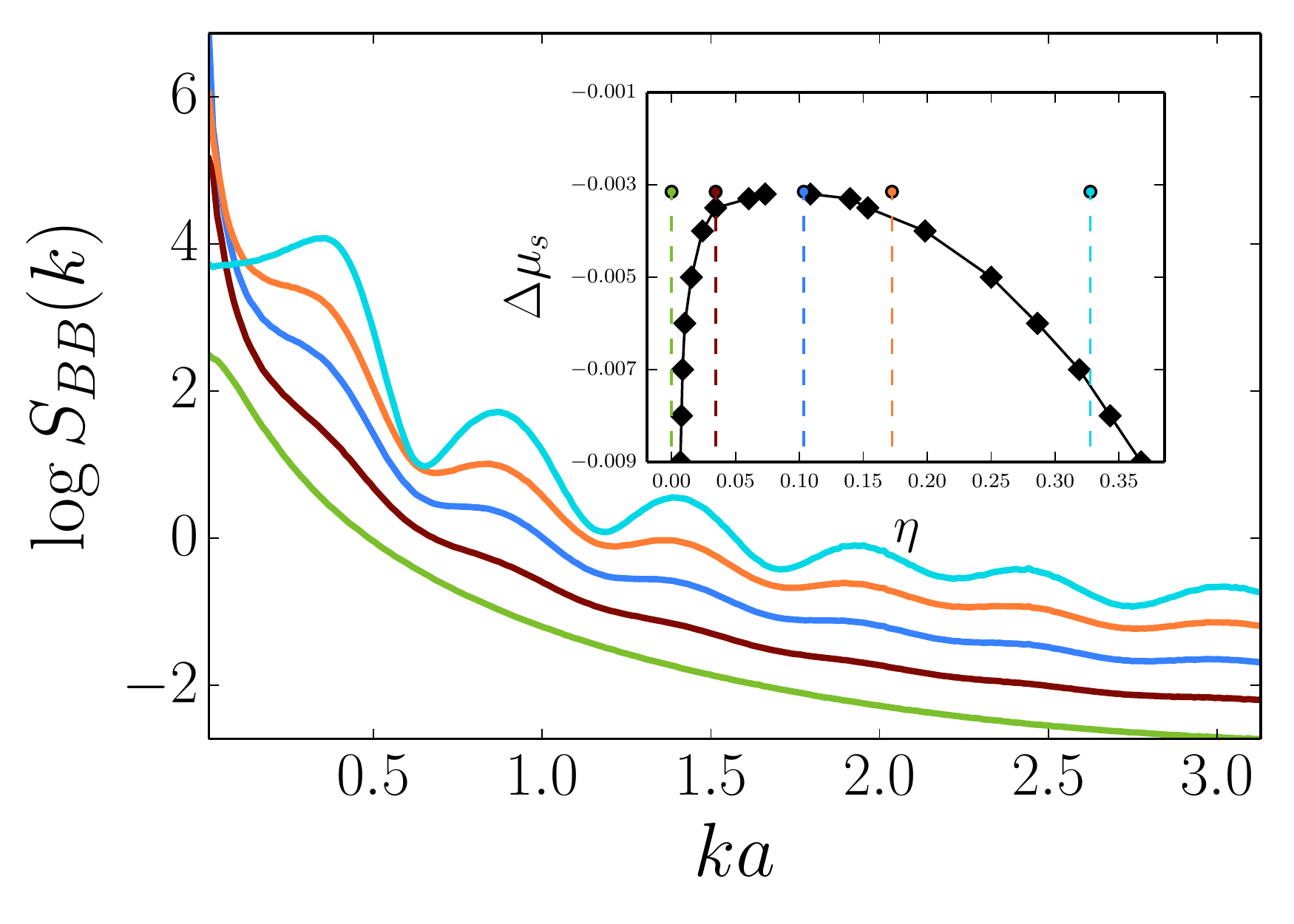}}
\subfigure[]{\includegraphics[width=3.0in]{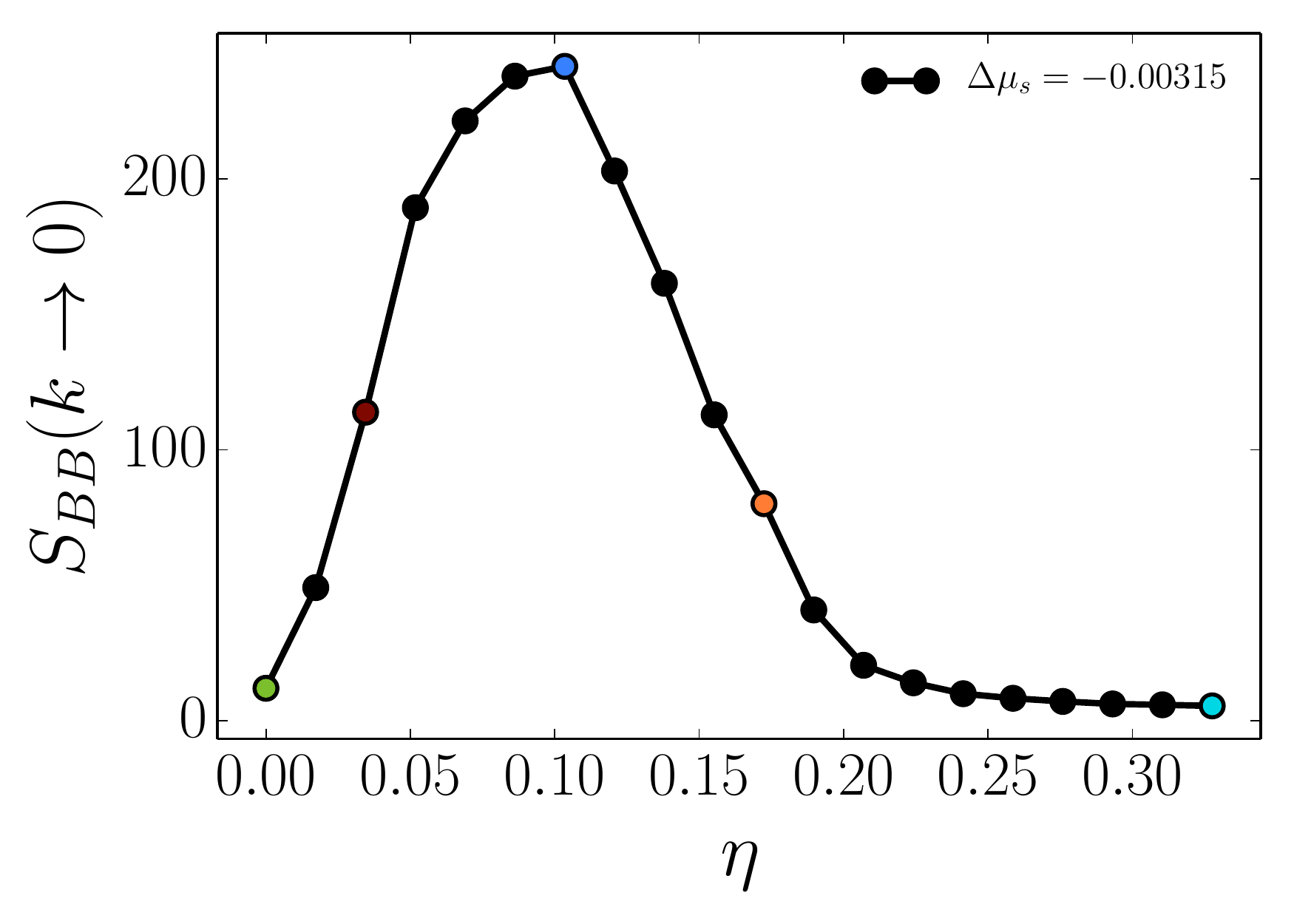}}
\caption{\label{SI_fig03ab} a) Partial structure factor $S_{BB}$ computed at $\tau = 0.025$, $\Delta \mu_s = -0.00315$, and different values of colloid packing fraction; $\eta = 0.0$ (green), $\eta = 0.0345$ (brown), $\eta = 0.0862$ (blue),$\eta = 0.1724$ (orange) and $\eta = 0.3276$ (cyan). The structure factors, were shifted by $0.5$ in log-scale for clarity. The inset shows the G-L binodal for $\tau = 0.025$ (black diamond symbols). b) The maximum value of the structure factor $S_{BB}(k\rightarrow0)$ vs. the packing fraction $\eta$ of the colloids.}
\end{figure}

At the G-L critical points, whose location can be gleaned from Fig. 3 b) of the paper the solvent-solvent (BB), colloid-colloid (CC) and solvent-colloid (BC) correlations decay with the same, diverging correlation length. Here in figure \ref{SI_fig03ab} a) we show the BB structure factor defined as $S_{BB}({k}) = (1/N) \left< n_{\textbf{k}} n_{-{\textbf{k}}} \right>$, where $n_{\textbf{k}}$, is the Fourier transform of the solvent occupancy profile \cite{Hansen2006}. We compute $S_{BB}({k})$ at a fixed temperature $\tau = 0.025$, and solvent chemical potential $\Delta \mu_s = -0.00315$, fixed very close to the critical value. We present results at several packing fractions of the colloid $\eta$, indicated by dots in the phase diagram, shown in the inset of Fig. \ref{SI_fig03ab} a). The long wavelength limits of the partial structure factors $S_{\alpha \beta}({k} = 0)$ diverge on approaching the critical point. In Fig. \ref{SI_fig03ab} b) we plot the limit $S_{BB}({k} = 0)$, obtained from a linear extrapolation of the simulation data, vs $\eta$, which shows a maximum corresponding to the state closest to the G-L critical point. Calculations of $S_{BB}({k} \rightarrow 0)$ vs $\eta$, close to the critical value, can yield a rough estimate of the G-L critical point. 

\begin{figure}
\centering
\subfigure[]{\includegraphics[width=3.0in]{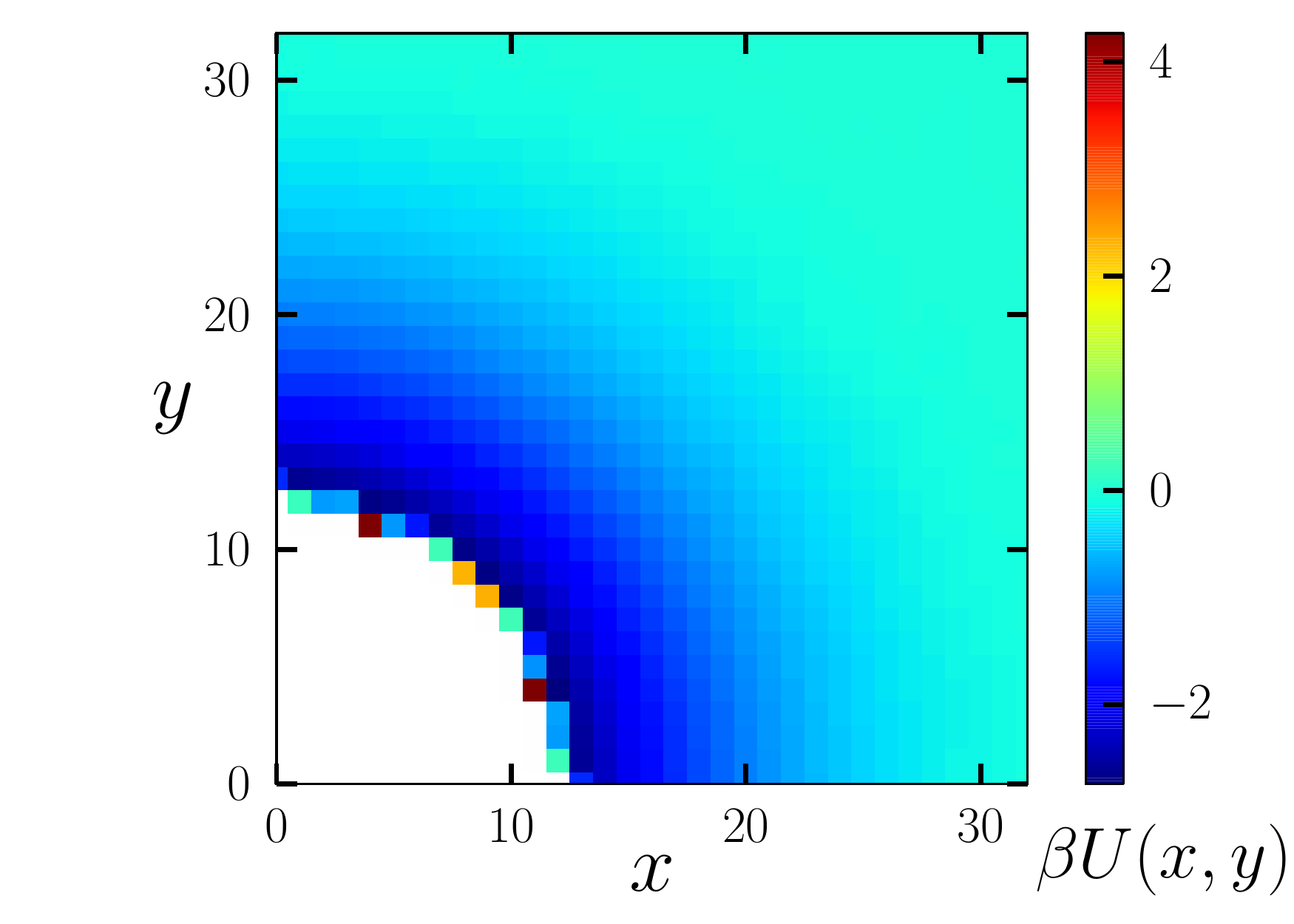}\label{SI_fig04_a}}
\subfigure[]{\includegraphics[width=3.0in]{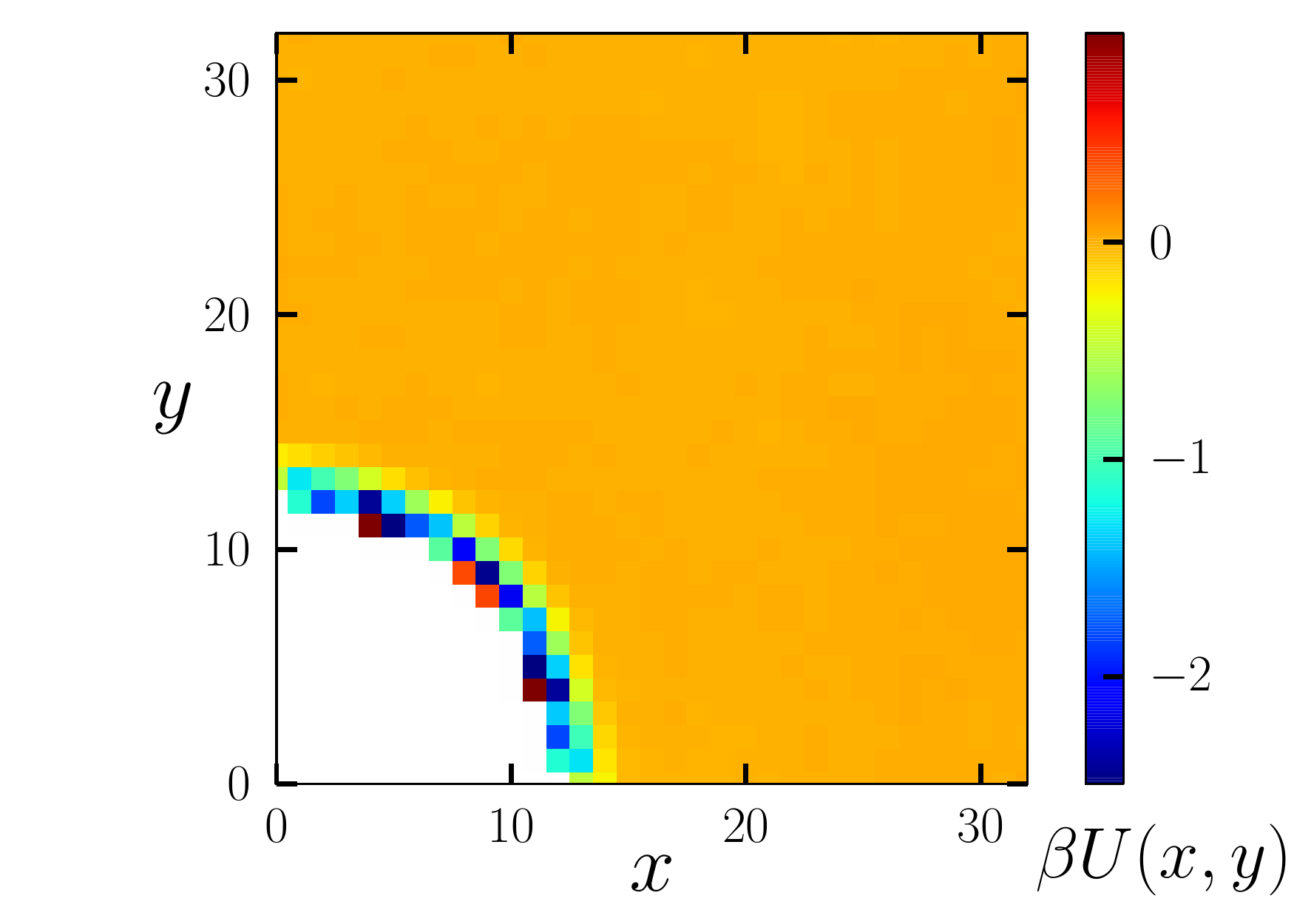}\label{SI_fig04_b}}
\caption{\label{SI_fig04} Effective two body potential between two colloids suspended in a solvent at $\tau = 0.05$ for a) $\Delta \mu_s = -0.01$ and b) $\Delta \mu_s = -0.5$. The area in white is inaccessible due to the hard core repulsion.}
\end{figure}

\subsection{Effective two-body interactions}

\begin{figure}
\centering
\includegraphics[scale=0.65,keepaspectratio]{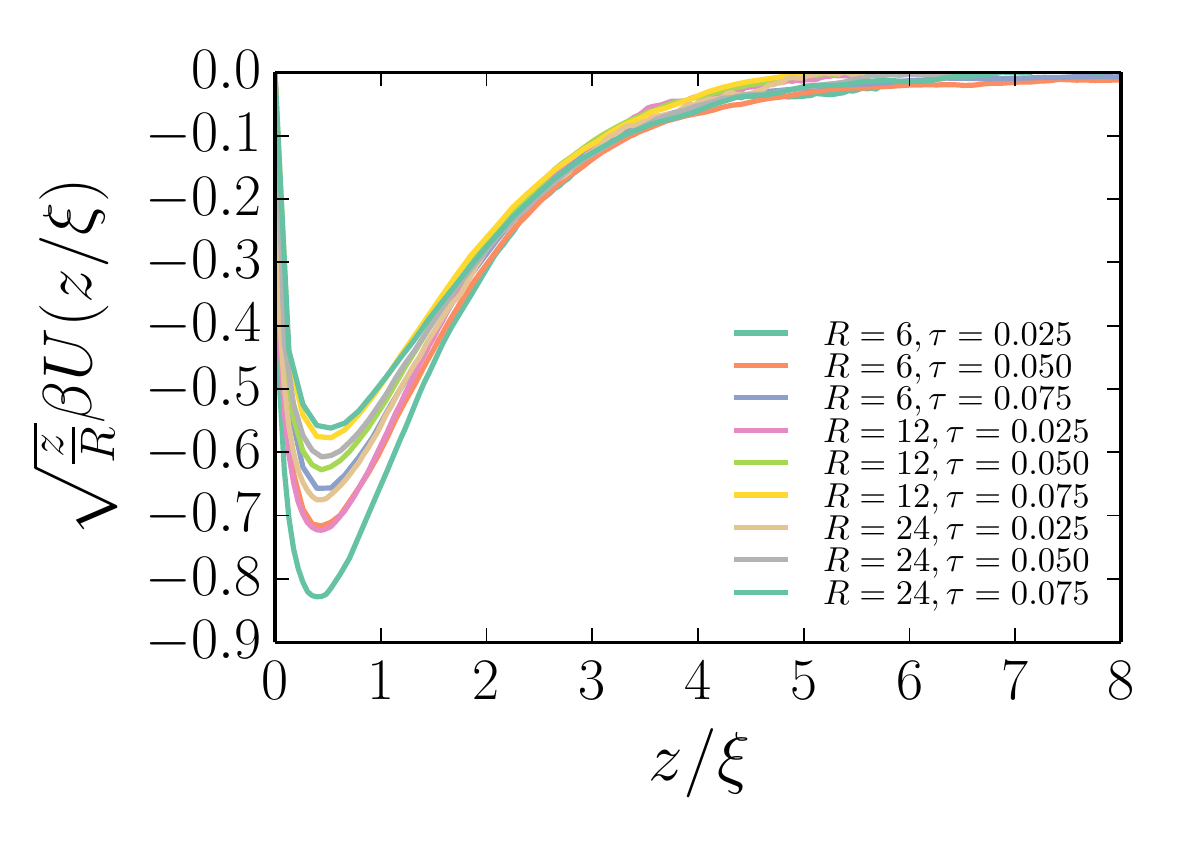}
\caption{\label{SI_fig05} Effective two-body interactions between a pair of colloids at $\Delta \mu_s = 0$, $\alpha=19$. $\xi$ is the correlation length of the solvent.}
\end{figure}

The effective two-body interactions were computed by simulating a system of two colloids at fixed $\{ \Delta \mu_s, \tau \}$. We fix the position of one colloid at the origin and measure the probability of finding the other at position $\{x,y\}$. We use the Transition Matrix Monte Carlo technique to make sure the colloids sample the entire range of distances $\{-L_{max}\leq x \leq L_{max}, -H_{max} \leq y \leq H_{max}\}$. The two body potential $U(x,y)$ is obtained as $U(x,y) = -kT \log(P(x,y)/P(\infty,\infty))$. 

The effective colloid-colloid interaction of the discretized colloids (refer to Fig. 1 in paper) in our lattice model is anisotropic; the strength of the interaction close to contact varies substantially. In figure \ref{SI_fig04} we plot the two-body potential measured at  $\Delta \mu_s = -0.01$ and $\Delta \mu_s = -0.5$ at  temperature $\tau = 0.05$, where it can be seen that lattice effects are pronounced when the range of the interaction is of the order of $1-3$ lattice sites. While these lattice effects play no role in G-L coexistence, they play a significant role in G-X coexistence. The crystal phase is facilitated by the colloids aligning along the more energetically favorable directions. 

The form of the effective colloid-colloid interaction between our colloidal discs depends on the proximity of the solvent reservoir to its critical point and, to some extent, on the value of the adsorption strength $\alpha$. In the scaling regime, i.e for small values of $\Delta \mu_s$ and $\tau$, the functional form of these effective interactions is known \cite{Zubaszewska2013, Machta2012,Burkhardt1995} theoretically. In Fig. \ref{SI_fig05} we plot the effective two-body interaction computed at $\Delta \mu_s = 0$ for colloids of different sizes at different temperatures. The distance between the colloids is scaled with the correlation length of the bulk reservoir. Our data shows good scaling behaviour, except at short distances where scaling is no longer applicable and where lattice effects become important. That we find good scaling gives us confidence that our simulations capture correctly the fluctuations responsible for the Casimir attraction.

\bibliography{casimir}

\end{document}